\documentclass[conference,a4paper]{IEEEtran}
\IEEEoverridecommandlockouts
\ifCLASSINFOpdf
  \usepackage[pdftex]{graphicx}
\else
\fi
\usepackage{url}
\usepackage[normalem]{ulem}

\newcommand{\fig}[4][width=120mm]{
  \begin{figure}[t]
    \centerline{\includegraphics[#1]{#4}}
    \caption{#2}
    \label{#3}
  \end{figure}
}

\newenvironment{tab}[4][]{
  \begin{table}[t]
    \caption{#2}
    \label{#3}
    #1
    \begin{center}
      \begin{tabular}{#4}
}{
      \end{tabular}
    \end{center}
  \end{table}
}

\newcommand{\mymin}{\mathop{\rm min}\limits}
\newcommand{\figref}[1]{\figurename\ \ref{#1}}
\newcommand{\tabref}[1]{\tablename\ \ref{#1}}
\newcommand{\eqnref}[1]{(\ref{#1})}
\newcommand{\Order}{\mathcal{O}}

\renewcommand\IEEEkeywordsname{Keywords}
\begin{document}
%
\title{Computing Information Quantity as Similarity Measure for Music Classification Task}

\author{\IEEEauthorblockN{Ayaka Takamoto, Mitsuo Yoshida, and Kyoji Umemura}
\IEEEauthorblockA{Department of Computer Science and Engineering\\
Toyohashi University of Technology \\
Toyohashi, Aichi, Japan\\
a153350@edu.tut.ac.jp, yoshida@cs.tut.ac.jp, umemura@tut.jp
}
\and
\IEEEauthorblockN{Yuko Ichikawa}
\IEEEauthorblockA{General Education Department,\\
National Institute of Technology, Tokyo College\\
Hachioji, Tokyo, Japan\\
yuko@tokyo-ct.ac.jp
}
} 

\maketitle
\IEEEpubid{\makebox[\columnwidth]{978-1-5386-3001-3/17/\$31.00~\copyright~2017 IEEE \hfill} \hspace{\columnsep}\makebox[\columnwidth]{ }}

\begin{abstract}

This paper proposes a novel method that can replace compression-based dissimilarity measure (CDM) in composer estimation task.  
The main features of the proposed method are clarity and scalability.  
First, since the proposed method is formalized by the information quantity, reproduction of the result is easier compared with the CDM method, where the result depends on a particular compression program.  
Second, the proposed method has a lower computational complexity in terms of the number of learning data compared with the CDM method.  
The number of correct results was compared with that of the CDM for the composer estimation task of five composers of 75 piano musical scores.  
The proposed method performed better than the CDM method that uses 
the file size compressed by a particular program.

\end{abstract}


\section{Introduction}

When people listen to music, they can determine many features, such as genre and composer. 
The genre of music is easy to determine without previous knowledge, but not the composer, even if you have some knowledge.  
The difficulty depends on what should be estimated.  
There are some existing studies for such estimation, which are based on machine learning ~\cite{Dannenberg1997, Sawada2000}.

The contribution of ~\cite{Sawada2000} implies that the feature that reflects the composer is short note sequence. 
Since the compression program is a kind of program which captures frequent sequences of data, it may not be suprising if we use a compression program to estimate composer. 
Actually, there is an interesting research ~\cite{Anan2012} that uses compression programs for composer estimation. 
They use the formular called NCD (Normalized Compression Distance). 
We focus on a similar but different similarity measure called compression-based dissimilarity measure (CDM)~\cite{Keogh2004}, which is tested in a wide range of data, not limited to music.
Both CDM and NCD are based on the same plinciple.
These principle are recently well presented in ~\cite{Louboutin2016}.
\IEEEpubidadjcol

Although a compression program is easy to use,
the result depends on the compression program and the behavior is
difficult to analyze.  Moreover, since the compression is
carried out with every known musical score, there is a concern
that the amount of calculation becomes enormous when we determine the
degree of similarity for a new musical score.  In this study, we propose a
novel method that is well formalized.  The proposed method
realized the scalability of a large number of learning data by
pre-processing the group of learning data.  Finally, the precision
of the proposed method was verified to be better than the method where the value of the CDM is
determined by the compressed file size.
\IEEEpubidadjcol

\section{Baseline Method}

In this section, we will describe baseline CDM method~\cite{Takamoto2016} for estimating the composers.
This work focuses on the improvement of CDM. 
They conducted experiments on a very simple system with CDM, but it still performs well for composer estimation task, in order to make the analysis of improvments possible.
We have followed this work because we are also interesting in CDM, although we are aiming at replacing CDM with the proposed method, rather than simply improving the CDM. 
In baseline method, the musical scores are first converted into string representation, where information for sound 'on' or 'off' is expressed. 
The string representation is a long sequence of character '0' for 'off' and character '1' for 'on'. The position of each character corresponds to a piano key number $key$ and timing number $time$, where $position = 88 \times time + key$. 
The number 88 is the number of keys on a  piano.
Second, it uses the CDM proposed by Keogh~\cite{Keogh2004} for a pair of musical scores.  
The CDM is defined as follows:

\begin{eqnarray}
{\rm CDM}(x,y) &=& \frac{C(xy)}{C(x)+C(y)} \label{eq_cdm}
\end{eqnarray}

where $C(x)$ is the compressed file size of string $x$, and $C(xy)$ is the compressed file size of the concatenation of $x$ and $y$.  
The value of the CDM shows the dissimilarity between the two strings.  
The more the patterns shared by the two strings, the smaller the CDM value of the two strings.  
It is based on the principle that the string has more similar patterns, such as repetitions, if the compressed file size of its concatenated string is smaller based on the assumption that if specific phrases of the composer exist, then he/she used them in other musical scores.  
The estimation of the composer is based on this concept. 
This method is based on the study in~\cite{Takamoto2016}.

It is interesting that the CDM, which is a simple function of a
compressed file size, can estimate the composer of musical scores.
However, there is an issue of scalability in the CDM. 
\figref{fig_comp_cdm} illustrates this issue, where $x$ is the string representation of a musical score of unknown composer, and $a_1$ to $a_{15}$ are musical scores of a composer $A$. 
The CDM is defined as a measure between two musical scores.  
In a previous study of composer estimation, an unknown musical score was compared with all the known musical scores, then the $k$-nearest neighbor method ($k$-NN) was applied~\cite{NLP-a} to the result.  
In general, when an application uses the relationship between two scores, an unknown musical score need to be compared with all the known musical scores.  
The larger is  the number of known scores, the more are the computation time  required for one new musical score.  
This method cannot be scaled up to a large number of known musical scores.

The study in~\cite{Takamoto2016} also argues that the compressed file size of string $x$
is the approximation of the information quantity. The study in~\cite{Takamoto2016} also
proposes to use offsetted compressed file size, where the value of the
offset is obtained by observing the behavior of a specific compression
program.  This method was reported to improve the number of correct
estimation significantly.  However, the problems of dependency on
the compression program and scalability remained the same with the CDM method.

\section{Proposed Method}

In this study, we formed a group of musical scores of the same
composer to address the scalability issue. Then, we computed the
information quantity based on the probability of substrings of a large
string. This large string corresponds to the group.
\figref{fig_comp_1} shows how the groups were used. As is in
\figref{fig_comp_cdm}, in \figref{fig_comp_1}, $x$ is the string
representation of a musical score of unknown composer, and $a_1$ to
$a_{15}$ are musical scores of a composer $A$.  The box shows that
these scores form a group, and there is one long string representation
for one group.  The information quantity is then computed using the
probability in $a_1 , a_2, ..., a_{15}$, and not the probability in
$x$.

Then, we computed the information quantity of an unknown musical
score using the method described in the next session.  The same process was
carried out for the musical scores of the other four composers.  We computed
the information quantity of the unknown musical score $x$ with the
group of each composer.  We obtained five information
quantities and determined that the composer of $x$ is the one whose
string had the least information quantity.

Using the pre-processing, the computation time of information quantity
of one music score does not depend on the number of music score
in a group. It only depends on the length of music score to judge.
Therefore, the number of computations for one unknown musical score
is proportional to the number of composers, rather than the
number of known musical scores.

\fig[width=0.65\columnwidth]{
Baseline system and other compression based approaches.
When a method uses the relationship between two scores, an unknown musical
score need to be compared with all the known musical scores.
}{fig_comp_cdm}{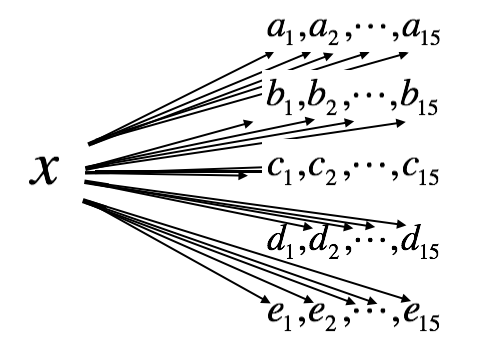}

\fig[width=0.65\columnwidth]{
Proposed system or scalable method.
By pre-prosessing through all music score in a group.
computation time of one music scores not depend the 
number of scores in group.
}{fig_comp_1}{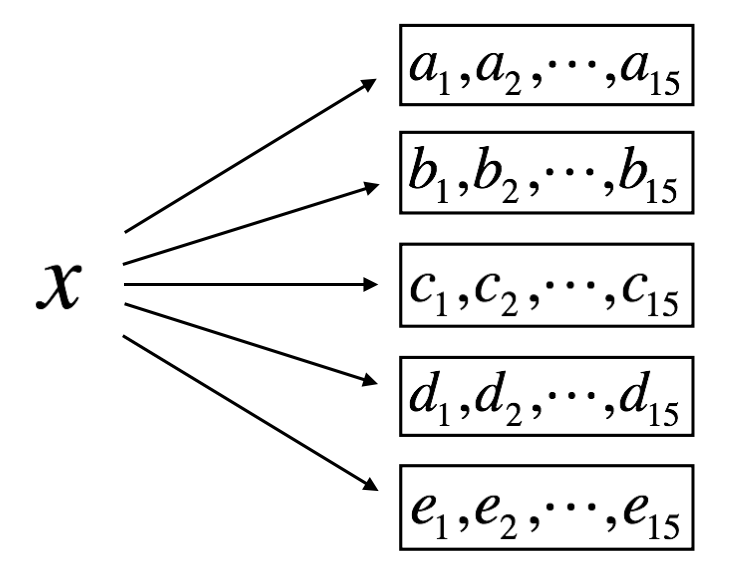}

\section{Information Quantity}

In general, we calculate the information quantity of a string from the
probabilities of characters in the string. However, we can make a good guess
that specified substrings, such as words emerge repeatedly in the real
strings. Therefore, in this study the calculation of the information
quantity of a string was performed using the emergent probabilities of all
substrings.

First, we consider the information quantity of one character.
In general, the information quantity of a certain event depends
on the occurring probability. Let the emergent probability of a
certain character $c$ be $P(c)$, then the information quantity of $c$
is expressed using self-information~\cite{NLP-b} as follows:

\begin{eqnarray}
I_c(c)=-\log_2P(c) \label{eq_logp}
\end{eqnarray}

Let us consider the information quantity for the case where we
treat the character sequence as a string.  Let the $i$-th
character of a string $S$ with length $N$ be $c_i$.  The character $c_i$
in $S$ is independent of each other.  The information quantity $I_c(S)$
of $S$ based on the characters is expressed as \eqnref{eq_str_info_c}.
The expression in \eqnref{eq_str_info_c} indicates that the string
information quantity is equal to the total sum of the information
quantity of the characters.

\begin{eqnarray}
  I_c(S) &=& - \log_2(\prod_{i = 1}^N P(c_i)) \nonumber\\
        &=& - \sum_{i = 1}^N \log_2 P(c_i)
\label{eq_str_info_c}
\end{eqnarray}

For the case of the string representation of a musical score,
specified substrings, such as motif may emerge repeatedly.  Thus, if we
assume that a string consists of some subsequences, then the information
quantity $I_s$ is expressed as \eqnref{eq_str_info_S}.

\begin{eqnarray}
I_s(S) = \mymin_{\pi_k \in \pi (S)} \Bigl( -\sum_{t \in \pi _k} \log_2P(t) \Bigr)
\label{eq_str_info_S}
\end{eqnarray}

where $\pi(S)$ is the set of all possible ways to divide $S$, which
includes $2^{N-1}$ ways, and $t$ is a member of divided strings (a substring).  More
precisely, we divide the strings into finer substrings and calculate
the information quantity as the sum of the information quantities of the new divided substrings.
The information quantity varies depending on the partition.  We should
take the minimum quantity because the more are the substrings considered, the
less becomes the information quantity of the string.  The number of
partitions is $2^{N-1}$, where $N$ is the length of the string.  Although
this is a large number, the minimum value is easily obtained in
$\Order(N^2)$, when a dynamic programming is used.

To implement a program that obtains $I_s(S)$, we require a module to compute
$P(t)$, where $t$ can be all substrings of the given large string.  An efficient data structure, called suffix array, can be used to obtain the frequency
of any substring~\cite{Manber1993}.  Using this data structure, whose
size is proportional to size of the large string, we can obtain the
frequency of a substring $t$ in the large string efficiently.
We used suffix array in the program implemented in this study.
There is also a more efficient data structure called suffix tree~\cite{Stringology}.  
Furthermore, there is a good algorithm that can construct
suffix tree in $\Order(N)$ time complexities, and $\Order(1)$ time
complexities to obtain the frequency of a substring using the suffix tree.
Maximum Likelihood Estimator (MLE) is usually used  to estimate the probability from the frequency.
We use MLE but we use $frequency - 1$ rather than $frequency$ in order to
make the computed value stable.

\section{Computational Complexity}

Let us examine by how much the computational complexity is
reduced by the proposed method compared with the existing method.

Let $l$ be the average length of a string representation of a musical
score. Let $c$ be the number of composers. Let $g$ be
the average number of musical scores in one group. Let $n$ be the
number of unknown musical scores.

First, the computational complexity to compress a string
representation of musical score is proportional to the length of the
string. Thus:

\begin{eqnarray}
   T_{compression} = \Order(l)
\end{eqnarray}

To estimate the composer of one musical score using CDM, we need to compute $g \times c$ compression:

\begin{eqnarray}
   T_{CDM-ONE} =  \Order(l \times g \times c))
\end{eqnarray}

When there are many musical scores, we need to repeat the above operation for each $n$ musical scores.

\begin{eqnarray}
   T_{CDM} = \Order(n \times l \times  g \times c)
\end{eqnarray}

To compute one information quantity in \figref{fig_comp_1}, we need to
compute two things: the pre-processing of the groups and 
to obtain the minimum of the considered partition.

\begin{eqnarray}
   T_{information-quantity} = \Order(g \times l + l^2)
\end{eqnarray}

To estimate the composer of one musical score using the proposed method, we need to compute the
information quantity $c$ times.

\begin{eqnarray}
   T_{Proposed-ONE} = \Order(c \times  g \times l + c \times l^2)
\end{eqnarray}

When there are many musical scores, we only require one pre-processing operation.
This is the reason why the proposed method is scalable.

\begin{eqnarray}
   T_{Proposed} = \Order(c \times  g \times l + n \times c \times l^2)
\end{eqnarray}

Both the complexity of $T_{CDM}$ and the $T_{Proposed}$ are proportional to
$c$. There is no difference with respect to $c$.  When $n$ is large, the
computational complexity of the proposed method is independent of $g$,
while that of the CDM is multiplied by $g$.  This means that when the
number of musical scores for each composer increases, the computational
complexity of the proposed method becomes smaller than CDM method.

The proposed method has a complexity that is proportional to the square of
the length of the unknown musical score, while that of the CDM method is
proportional to the length.  This is because the proposed method considers
all substrings, while the compression program only considers some subsets
of the substrings.  As a result, the proposed method requires a large value of $g$ when
the $l$ is large.

\section{Evaluation}

For the evaluation, the result can be much better than it should be if we include the same
musical score as $x$ in some of the known musical scores. Therefore, we
have to change our setting from \figref{fig_comp_1} into
\figref{fig_comp_2} to measure the correctness of the methods.  The
musical score in question should be intentionally excluded from the
set of known musical scores.  In the CDM method, the
CDM between the same musical scores was not computed using the one-leave-out method. \figref{fig_comp_2} corresponds to this approach. In doing so, we
need to pre-process many times, and this is only for the evaluation,
and not the actual estimation.

In \figref{fig_comp_2}, A, B, C, D, and E indicate the composers and $a_1,
\cdots, a_{15}$ denote the musical scores of composer A.  When we need to
estimate the composer of $a_1$, we remove $a_1$ from the group of
composer A, and create a new group data of the remaining musical scores. Then, we
calculate the information quantities of $a_1$ with each of the five
grouping data. We estimate that the composer of $a_1$ is the one whose
string attains the least information quantity.
Then, we determine the estimation from the information that the composer of $a_1$ is A.
This information is used
only for determining the correctness of the estimation.

\fig[width=0.65\columnwidth]{
For the evaluation, the result can be much better than it should be if we include the same
musical score in some of the known musical scores. Therefore,
the musical score in question should be intentionally excluded from the
set of known musical scores for evaluation. This corresponds to one-leave-out method.
}{fig_comp_2}{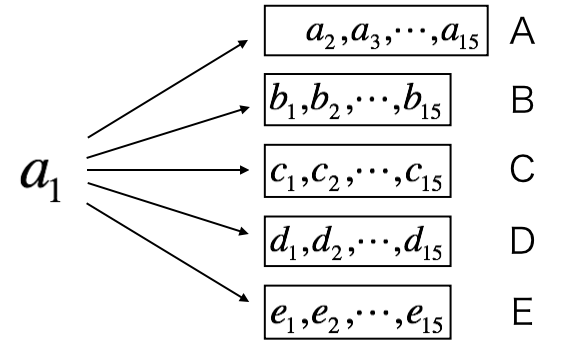}


A summary of the total correct results is presented in \tabref{tab_summary}.
In the estimations of 75 musical scores, the proposed method yielded
55 correct results.  Since the task was to select one composer out of
five composers, a random choice can achieve 20\% correct answers.  Our method achieved more than 70\% correct answers.  This suggests that
the proposed method can estimate the composer.
Unlike the CDM, the proposed method is formalized as the estimation of
information quantity, and is not dependent on a particular compression
program.  Therefore, reproduction of the result should be much easier than
the CDM.

\begin{tab}{Summary}{tab_summary}{cc|ccc}
& & \multicolumn{3}{c}{System} \\
& & Proposed & CDM & Offseted CDM \\\hline
& Bach    & 9 & 10 & 11 \\
& Chopin  & 9 & 5 & 6 \\
Group & Debussy & 14 & 11 & 12 \\
& Mozart  & 13 & 7 & 11 \\
& Satie   & 10 & 8 & 8 \\\hline
& Total  & 55 & 41 & 48
\end{tab}


As presented in \tabref{tab_summary}, the proposed method
yielded more correct results than previous methods. We performed the McNemar's 
test of the proposed method  with the original CDM and with offsetted CDM~\cite{Takamoto2016}. As presented in
table \tabref{tab_mcnemar_CDM}, the proposed method performed better than the CDM
with a significance of $\alpha< 0.01$, although we could not achieve the
statistical significance in \tabref{tab_mcnemar_offset}.

Since offsetted CDM~\cite{Takamoto2016} aimed to obtain a more precise
value of the information quantity rather than the compression, the behavior of the
proposed method would be similar to offsetted CDM~\cite{Takamoto2016}. However, it can be seen that the proposed method is
independent of the implementation of a particular compression program,
while ~\cite{Takamoto2016} depends on a compression program, bzip2.

\tabref{tab_bach} to \tabref{tab_satie} present the detailed results of
applying the proposed method to 15 pieces for each five composer:
Bach, Chopin, Debussy, Mozart, and Satie.  Each column below the label
``Music'', contains the identifier starting with its composer and
ending with the identification number.  The column of the name of each
composer contains information quantities using the group.  The value
is truncated to the nearest integer.  The case where the information quantity of a
given score is the least, it is underlined.  The corresponding
composer of the underlined value is the estimation using the proposed method.
The column ``Result'' indicates whether the estimation is correct or not,
where ``1'' is correct, and ``0'' is incorrect.  The column ``CDM'' is
the result of using the baseline method, which follows the technique in
~\cite{Keogh2004}, where $C(x)$ is the file size of the compressed
file using bzip2.  The column ``offset'' is the result from a previous
study~\cite{Takamoto2016},  where
$C(x)$ is the offsetted size of the compressed file.


\begin{tab}{McNemar's Test Between Proposed Method and CDM without Offset}{tab_mcnemar_CDM}{cc|cc|c}
& & \multicolumn{3}{c}{CDM} \\
& & Correct result & Incorrect result & Total \\\hline
Proposed & Correct result & 38 & 17 & 55 \\
& Incorrect result   & 3 & 17 & 20 \\\hline
& Total  & 41 & 34 & 75
\end{tab}

\begin{tab}{McNemar's Test Between Proposed Method and CDM with Offset}{tab_mcnemar_offset}{cc|cc|c}
& & \multicolumn{3}{c}{Offseted CDM} \\
& & Correct result & Incorrect result & Total \\\hline
Proposed & Correct result & 43 & 12 & 55 \\
& Incorrect result   & 5 & 15 & 20 \\\hline
& Total  & 48 & 27 & 75
\end{tab}

\section{Discussion}

Sometimes, the methods of smaller computational complexities may be slower
in actual number of data when the data is not big enough.  Currentlly, this is the case of the proposed method.
Since the current group consists of 15 musical scores, the proposed method was
slower than the CDM method in the current condition. There are several
reasons for this inefficiency.  The most important reason is that the 
proposed method requires a computation time that is proportional to the
square of the length of the string, while the CDM (or compression program)
requires a computational time that is proportional to the length.
Furthermore, the string representation usually consists of
more than 10000 characters.  This length could be the reason
for the inefficiency.

We may improve the computation time by limiting the set of substring that
is used to compute the information quantity and concentrate computation
resources to the string that should be effective.  With respect to the
efficiency of the compression program, the compression program may not
consider all the strings.  Thus, we need a heuristic technique that may be
common to the compression program.

There may be other viewpoints in terms of the contributions of this work.
The application of the CDM is not limited to this task, but also applies to various
types of tasks.  We may search an appropriate task where there are many
samples in a class and the length of data to consider is small.

There is a method to calculate information quantity for 
estimation similarities called normalized compression distance
(NCD)~\cite{Cilibrasi2005}.  Some studies have applied this method in the field of
biological information~\cite{Li2003} and in the field of musical
information~\cite{Cilibrasi2004, Ahonen2011}.  The order of
computational complexity with this method is the same as in the CDM.

It seems useful to select data patterns or subsequences that
emerged repeatedly from known data and improve the compression
program so that the information quantity of an unknown data is
calculated with the selected data.  However, some compression
programs have a limit in the number of words registered in their
dictionary.  Since our proposed method considers all the substrings
of the given string, the method does not suffer this limitation, and we
may state that it uses a larger dictionary compared with any other method.

\section{Conclusion}

We proposed a novel method that can replace the CDM  method for
the composer estimation task.  The main feature of the proposed method is
the pre-processing of the grouped data of each composer.  We showed that the
computational complexity in terms of the number of known musical scores was
smaller than in the CDM.  This means that the proposed method is
scalable.  We also verified that the number of correct estimations obtained was 55 out of 75 estimations.  This result is better
than the estimation result of the CDM method.  Moreover, the computational
complexity to determine a new score was smaller than the CDM method.  Based on
the number of correct results and the order of computational
complexity, we can conclude that computing the information quantity with grouping is
effective.

\bibliographystyle{IEEEtran}
\bibliography{icaicta2017}

\begin{thebibliography}{10}
\providecommand{\url}[1]{#1}
\csname url@samestyle\endcsname
\providecommand{\newblock}{\relax}
\providecommand{\bibinfo}[2]{#2}
\providecommand{\BIBentrySTDinterwordspacing}{\spaceskip=0pt\relax}
\providecommand{\BIBentryALTinterwordstretchfactor}{4}
\providecommand{\BIBentryALTinterwordspacing}{\spaceskip=\fontdimen2\font plus
\BIBentryALTinterwordstretchfactor\fontdimen3\font minus
  \fontdimen4\font\relax}
\providecommand{\BIBforeignlanguage}[2]{{%
\expandafter\ifx\csname l@#1\endcsname\relax
\typeout{** WARNING: IEEEtran.bst: No hyphenation pattern has been}%
\typeout{** loaded for the language `#1'. Using the pattern for}%
\typeout{** the default language instead.}%
\else
\language=\csname l@#1\endcsname
\fi
#2}}
\providecommand{\BIBdecl}{\relax}
\BIBdecl

\bibitem{Dannenberg1997}
R.~B. Dannenberg, B.~Thom, and D.~Watson, ``A machine learning approach to
  musical style recognition,'' in \emph{Proceedings of International Computer
  Music Conference}, 1997, pp. 344--347.

\bibitem{Sawada2000}
T.~Sawada and K.~Satoh, ``Composer classification based on patterns of short
  note sequences,'' in \emph{Proceedings of the AAAI-2000 Workshop on AI and
  Music}, 2000, pp. 24 -- 27.

\bibitem{Anan2012}
Y.~Anan, K.~Hatano, H.~Bannai, M.~Takeda, and K.~Satoh, ``Polyphonic music
  classification on symbolic data using dissimilarity functions.'' in
  \emph{ISMIR}, 2012, pp. 229--234.

\bibitem{Keogh2004}
E.~Keogh, S.~Lonardi, and C.~A. Ratanamahatana, ``Towards parameter-free data
  mining,'' in \emph{Proceedings of the Tenth ACM SIGKDD International
  Conference on Knowledge Discovery and Data Mining}, ser. KDD '04.\hskip 1em
  plus 0.5em minus 0.4em\relax New York, NY, USA: ACM, 2004, pp. 206--215.

\bibitem{Louboutin2016}
C.~Louboutin and D.~Meredith, ``Using general-purpose compression algorithms
  for music analysis,'' \emph{Journal of New Music Research}, 2016.

\bibitem{Takamoto2016}
A.~Takamoto, M.~Umemura, M.~Yoshida, and K.~Umemura, ``Improving compression
  based dissimilarity measure for music score analysis,'' in \emph{Proceedings
  of 2016 International Conference On Advanced Informatics: Concepts, Theory
  And Application (ICAICTA)}, Aug 2016, pp. 1--5.

\bibitem{NLP-a}
C.~D. Manning, H.~Sch{\"u}tze \emph{et~al.}, \emph{Foundations of statistical
  natural language processing}.\hskip 1em plus 0.5em minus 0.4em\relax MIT
  Press, 1999, vol. 999, pp. 604--606.

\bibitem{NLP-b}
------, \emph{Foundations of statistical natural language processing}.\hskip
  1em plus 0.5em minus 0.4em\relax MIT Press, 1999, vol. 999, pp. 61--63.

\bibitem{Manber1993}
U.~Manber and G.~Myers, ``Suffix arrays: A new method for on-line string
  searches,'' \emph{SIAM Journal on Computing}, vol.~22, no.~5, pp. 935--948,
  1993.

\bibitem{Stringology}
M.~Crochemore and W.~Rytter, \emph{Jewels of stringology: text
  algorithms}.\hskip 1em plus 0.5em minus 0.4em\relax World Scientific, 2003,
  pp. 91--95.

\bibitem{Cilibrasi2005}
R.~Cilibrasi and P.~M.~B. Vitanyi, ``Clustering by compression,'' \emph{IEEE
  Transactions on Information Theory}, vol.~51, no.~4, pp. 1523--1545, April
  2005.

\bibitem{Li2003}
M.~Li, X.~Chen, X.~Li, B.~Ma, and P.~Vit\'{a}nyi, ``The similarity metric,'' in
  \emph{Proceedings of the Fourteenth Annual ACM-SIAM Symposium on Discrete
  Algorithms}, ser. SODA '03.\hskip 1em plus 0.5em minus 0.4em\relax
  Philadelphia, PA, USA: Society for Industrial and Applied Mathematics, 2003,
  pp. 863--872.

\bibitem{Cilibrasi2004}
R.~Cilibrasi, P.~Vit{\'a}nyi, and R.~De~Wolf, ``Algorithmic clustering of music
  based on string compression,'' \emph{Computer Music Journal}, vol.~28, no.~4,
  pp. 49--67, 2004.

\bibitem{Ahonen2011}
T.~E. Ahonen, K.~Lemstr{\"o}m, and S.~Linkola, ``Compression-based similarity
  measures in symbolic, polyphonic music.'' in \emph{Proceesings of ISMIR2011},
  2011, pp. 91--96.

\end{thebibliography}

\begin{tab}[\fontsize{6pt}{9pt}\selectfont]{Results of Bach's musical scores}{tab_bach}{c|ccccc|c||c|c}
Music  & Bach & Chopin & Debussy & Mozart & Satie & Result & CDM & Offset \\\hline
Bach01 & 27451 & 24371 & \uline{23512} & 25252 & 23938 & 0 & 0 &0\\
Bach02 & 7444 & 6819 & 7004 & \uline{6352} & 7574 & 0 &0&0\\
Bach03 & 22101 & 18742 & 18379 & 19855 & \uline{17964} & 0& 0&0\\
Bach04 & \uline{2711} & 3376 & 3509 & 2846 & 3464 & 1 &1&1\\
Bach05 & \uline{3093} & 3847 & 3827 & 3515 & 3908 & 1 &1&1\\
Bach06 & 3128 & 3149 & 3491 & \uline{2827} & 3420 & 0 &0&1\\
Bach07 & \uline{4796} & 6301 & 6487 & 5875 & 6740 & 1 &1&1\\
Bach08 & \uline{5278} & 5756 & 6017 & 5585 & 6018 & 1 &1&1\\
Bach09 & \uline{4068} & 4159 & 4239 & 4174 & 4468 & 1 &1&1\\
Bach10 & 5817 & 6051 & 5717 & \uline{5622} & 5977 & 0 &0&0\\
Bach11 & \uline{3411} & 4115 & 4171 & 3941 & 4380 & 1 &1&1\\
Bach12 & \uline{2847} & 3383 & 3444 & 3079 & 3592 & 1 &1&1\\
Bach13 & \uline{2577} & 3020 & 3163 & 2874 & 3221 & 1 &1&1\\
Bach14 & \uline{4736} & 5039 & 5185 & 4767 & 5173 & 1 &1&1\\
Bach15 & 2943 & 3065 & 3170 & \uline{2910} & 3197 & 0 &1&1\\
\hline
Total &&&&&&9&10&11
\end{tab}

\begin{tab}[\fontsize{6pt}{9pt}\selectfont]{Results of Chopin's musical scores}{tab_chopin}{c|ccccc|c||c|c}
Music  & Bach & Chopin & Debussy & Mozart & Satie & Result &CDM & Offset\\\hline
Chopin01 & 19541 & 17771 & \uline{17584} & 17969 & 18231 & 0 &0&0\\
Chopin02 & 15815 & 15421 & 14967 & \uline{14930} & 15409 & 0 &0&0\\
Chopin03 & 9114 & \uline{8287} & 8533 & 8891 & 8541 & 1 &0&0\\
Chopin04 & 21942 & 21665 & 21541 & 23272 & \uline{21159} & 0 &0&1\\
Chopin05 & 9863 & \uline{8631} & 9470 & 9094& 9058 & 1 &1&1\\
Chopin06 & 13492 & \uline{13032} & 13408 & 13096 & 13624 & 1 &0&0\\
Chopin07 & 65530 & \uline{54654} & 58969 & 59320 & 58647 & 1 &1&1\\
Chopin08 & 68263 & 61142 & 59976 & 67106 & \uline{60229} & 0 &0&0\\
Chopin09 & 14961 & \uline{9395} & 13513 & 12461 & 12704 & 1 &1&1\\
Chopin10 & 19985 & \uline{13767} & 17612 & 17426 & 17165 & 1 &1&1\\
Chopin11 & 20405 & \uline{17357} & 19401 & 18954 & 17931 & 1 &0&0\\
Chopin12 & 24312 & \uline{20111} & 21933 & 22378 & 20811 & 1 &0&1\\
Chopin13 & 18227 & 15593 & 15819 & 16123 & \uline{15566} & 0 &0&0\\
Chopin14 & 27187 & 23606 & 23634 & \uline{23253} & 24323 & 0 &0&0\\
Chopin15 & 10584 & \uline{9210} & 9584 & 9619 & 9430 & 1&1&0\\
\hline
Total &&&&&& 9&5&6
\end{tab}

\begin{tab}[\fontsize{6pt}{9pt}\selectfont]{Results of Debussy's musical scores}{tab_debussy}{c|ccccc|c||c|c}
Music  & Bach & Chopin & Debussy & Mozart & Satie & Result &CDM&Offset\\\hline
Debussy01 & 9354 & 7794 & \uline{7112} & 9766 & 7236 & 1 &1&1\\
Debussy02 & 26680 & 23481 & \uline{22086} & 24583 & 24005 & 1 &1&1\\
Debussy03 & 18945 & 17512 & \uline{16432} & 18015 & 17533 & 1 &1&1\\
Debussy04 & 8063 & 7009 & \uline{6685} & 7759 & 6732 & 1 &1&1\\
Debussy05 & 61477 & 58790 & \uline{51872} & 68706 & 55066 & 1 &0&0\\
Debussy06 & 10747 & 10289 & \uline{8930} & 10398 & 9358 & 1 &1&1\\
Debussy07 & 6248 & 5567 & \uline{4876} & 5177 & 4946 & 1 &0&1\\
Debussy08 & 37096 & 33933 & \uline{30807} & 36659 & 34117 & 1& 1&1\\
Debussy09 & 27645 & 25809 & \uline{22510} & 28316 & 24011 & 1 &1&1\\
Debussy10 & 24904 & 22108& \uline{20628} & 23234 & 21491 & 1 &1&1\\
Debussy11 & 19554& 18317 & \uline{17215} & 19764 & 17722 & 1 &1&1\\
Debussy12 & 26298 & 23519 & \uline{20882} & 26313 & 23024 & 1 &1&1\\
Debussy13 & 14808 & 14524 & \uline{13286} & 13942 & 14126 & 1 &1&1\\
Debussy14 & 12767 & 11919 & \uline{10769} & 11536 & 11327 & 1 &0&0\\
Debussy15 & 11136 & 11259 & 10988 & \uline{10904} & 11387 & 0&0&0\\
\hline
Total &&&&&&14&11&12
\end{tab}

\begin{tab}[\fontsize{6pt}{9pt}\selectfont]{Results of Mozart's musical scores}{tab_morzart}{c|ccccc|c||c|c}
Music  & Bach & Chopin & Debussy & Mozart & Satie & Result & CDM& Offset\\\hline
Mozart01 & 10249 & 8417 & \uline{7873} & 9208 & 7992 & 0& 0&0\\
Mozart02 & 14406 & 12283 & 13247 & \uline{11508} & 12686 & 1&1 &1\\
Mozart03 & 4010 & 3814& 3954 & \uline{3199} & 3801 & 1 &0&1\\
Mozart04 & 7297 & 6862 & 7216& \uline{6730} & 7119 & 1 &1&1\\
Mozart05 & 15086 & 12929 & 13605 & \uline{11268} & 14960 & 1 &1&1\\
Mozart06 & 36692 & \uline{34726} & 35098 & 35133 & 37189 & 0 &1&1\\
Mozart07 & 2011 & 1867 & 1987 & \uline{1504} & 1917 & 1 &0&0\\
Mozart08 & 4121 & 3982 & 3815 & \uline{3541} & 4399 & 1 &0&1\\
Mozart09 & 6537& 6194 & 6392 & \uline{5335} & 6679 & 1 &0&0\\
Mozart10 & 2635 & 2506 & 1999 & \uline{1883} & 2102 & 1 &0&0\\
Mozart11 & 23620 & 19827 & 22406 & \uline{19113} & 23005 & 1 &1&1\\
Mozart12 & 8347 & 8277 & 8181 & \uline{5915} & 6873 & 1 &0&1\\
Mozart13 & 12219 & 12680 & 13134 & \uline{11361} & 13393 & 1 &0&1\\
Mozart14 & 11809 & 11318 & 11548 & \uline{10456} & 11621& 1 &1&1\\
Mozart15 & 52876 & 46259 & 48003 & \uline{43937} & 49226 & 1 &1&1\\
\hline
Total &&&&&&13&7&11
\end{tab}

\begin{tab}[\fontsize{6pt}{9pt}\selectfont]{Results of Satie's musical scores}{tab_satie}{c|ccccc|c||c|c}
Music  & Bach & Chopin & Debussy & Mozart & Satie & Result &CDM&Offset\\\hline
Satie01 & 2195 & 2062 & 2043 & \uline{1631} & 2216 & 0& 0&0\\
Satie02 & 23498 & 19024 & 19447 & 23665 & \uline{18218} & 1 &0&0\\
Satie03 & 17168 & 21461 & 18440 & 24132 & \uline{4917} & 1 &1&1\\
Satie04 & 7182 & 7522& 7746 & 8586 & \uline{4761} & 1 &1&1\\
Satie05 & 12186 & 14146 & 11912 & 15434 & \uline{4347} & 1 &1&1\\
Satie06 & 42936& 32240 & 32877 & 37414 & \uline{24716} & 1 &0&0\\
Satie07 & 43494 & 34264 & 36157 & 39355 & \uline{28590} & 1 &0&0\\
Satie08 & 13254 & 10550 & \uline{8655} & 10950 & 9464 & 0 &0&0\\
Satie09 & 4549 & 4489 & \uline{4036} & 4420 & 4301 & 0 &0&0\\
Satie10 & 17845 & 13940 & 13761 & 16366 & \uline{9969} & 1 &1&1\\
Satie11 & 1242 & 1233 & 1240 & \uline{935} & 1071 & 0 &0&0\\
Satie12 & 14957 & 14223 & 13550 & 15689 & \uline{10827} & 1 &1&1\\
Satie13 & 12061 & 11894 & 10876 & 12818 & \uline{8932} & 1 &1&1\\
Satie14 & 10464 & 10299 & 9578 & 11628 & \uline{6917} & 1 &1&1\\
Satie15 & 7030 & 6701& \uline{5432} & 8270 & 5845 & 0&1&1\\
\hline
Total &&&&&& 10&8&8
\end{tab}

\end{document}